\documentclass[useAMS,usenatbib]{mn2e}
\usepackage{graphicx}    
\usepackage{dcolumn}     
\usepackage{bm}          
\usepackage{amssymb,amsmath}  
\usepackage{color}
\usepackage{fixltx2e}
\usepackage{lastpage}
\usepackage{xfrac, multirow, url, balance, placeins}

\newcommand{\be}{\begin{equation}}
\newcommand{\ee}{\end{equation}}
\newcommand{\bea}{\begin{eqnarray}}
\newcommand{\eea}{\end{eqnarray}}

\usepackage[stable]{footmisc}

\title[Cosmological performance of future SKA HI galaxy surveys]{Cosmological performance of SKA HI galaxy surveys}

\author[S. Yahya, P. Bull, M. G. Santos, M. Silva, R. Maartens, P. Okouma and B. Bassett ]{S. Yahya$^{1}$\footnotemark[0]\thanks{Email: sahbayahya@gmail.com},  
P. Bull$^2$, Mario G. Santos$^{1,3,4}$, M. Silva$^{1,4}$, R. Maartens$^{1,5}$,
 \newauthor  
P. Okouma$^{1,6,7}$ and B. Bassett$^{6,7,8}$
\\\\\\
$^{1}$Physics Department, University of the Western Cape, Cape Town 7535, South Africa\\
$^2$Institute of Theoretical Astrophysics, University of Oslo, P.O.\ Box 1029 Blindern, N-0315 Oslo, Norway \\
$^{3}$SKA South Africa, The Park, Park Road, Cape Town 7405, South Africa\\
$^{4}$CENTRA, Instituto Superior Tecnico, Technical University of Lisbon, Lisboa 104900, Portugal\\
$^5$Institute of Cosmology \& Gravitation, University of Portsmouth, Portsmouth PO1 3FX, United Kingdom\\
$^{6}$African Institute for Mathematical Sciences, Cape Town 7945, South Africa\\
$^{7}$Department of Mathematics \& Applied Mathematics, University of Cape Town, Cape Town 7701, South Africa\\
$^{8}$South African Astronomical Observatory, Cape Town, South Africa
}

\begin{document}

\date{\today}
\pagerange{\pageref{firstpage}--\pageref{LastPage}} \pubyear{2014}
\maketitle

\label{firstpage}
\begin{abstract}
The Square Kilometre Array (SKA) will conduct the biggest spectroscopic galaxy survey ever, by detecting the 21cm emission line of neutral hydrogen (HI) from around a billion galaxies over $\sfrac{3}{4}$ of the sky, out to a redshift of $z \sim 2$. This will allow the redshift-space matter power spectrum, and corresponding dark energy observables, to be measured with unprecedented precision.
In this paper, we present an improved model of the HI galaxy number counts and bias from semi-analytic simulations, and use it to calculate the expected yield of HI galaxies from surveys with a variety of Phase 1 and 2 SKA configurations. We illustrate the relative performance of the different surveys by forecasting errors on the radial and transverse scales of the baryon acoustic oscillation (BAO) feature, finding that the full ``billion galaxy survey'' with SKA2 will deliver the largest dark energy figure of merit of any current or future large-scale structure survey.

\end{abstract}

\begin{keywords}
cosmology : radio surveys --- galaxy power spectrum --- baryonic oscillations
--- dark energy
\end{keywords}

\section{Introduction}

The Square Kilometre Array (SKA) is a giant radio telescope array, to be constructed across two sites, in South Africa and Western Australia. The first phase is due for completion in 2020, with a second phase (with about ten times the sensitivity and twenty times the field of view) planned for 2025. One of the key science aims of the SKA is to probe the nature of dark energy by mapping out large-scale structure, primarily using the 21cm emission line of neutral hydrogen (HI) to detect galaxies and measure their redshifts with high (spectroscopic) precision.

At present, HI galaxy surveys (e.g. HIPASS, \citealt{2004MNRAS.350.1195M}) are quite small compared to optical and near-infrared counterparts like BOSS and WiggleZ, limiting their use for precision cosmology. The unprecedented sensitivity and field of view of the SKA will allow for dramatically faster survey speeds, making it possible to map the galaxy distribution out to high redshifts over most of the sky. The end result will be sample variance-limited observations over a truly gigantic survey volume, allowing HI surveys to outperform other methods in terms of precision cosmological constraints, and making it possible to probe ultra-large scales and novel wide-angle effects \citep{Abdalla:2009wr, 2014arXiv1409.8286C}.

The current best cosmological constraints from large-scale structure surveys come from observations of the baryon acoustic oscillations (BAO). The BAO feature is a preferred clustering scale imprinted in the matter distribution by acoustic oscillations in the coupled photon-baryon fluid around the time of decoupling \citep{Bassett:2009mm}. The radial and transverse BAO scales depend on the Hubble rate, $H(z)$, and the angular diameter distance, $D_A(z)$, as well as the (comoving) sound horizon in the `baryon drag' epoch, $r_s(z_d)$. The comoving sizes of the BAO feature along and across the line of sight are given by
\begin{equation}\label{DandH}
s_\Vert(z) = \frac{c \Delta z}{H(z)}\,,  \quad s_\perp(z) = (1+z) D_{A}(z) \Delta \theta,
\end{equation}
where the redshift extent $\Delta z$ and angular size $\Delta \theta$ of the BAO feature in the galaxy correlation function are the observables. In the absence of redshift-space distortions (RSDs) and  nonlinear effects, we have $s_\perp = s_\parallel = r_s(z_d)$, which can be precisely estimated from (e.g.) CMB measurements. The BAO scale is therefore a `standard ruler', with which we can obtain precise constraints on $D_A$ and $H$, and thus the dark energy equation of state, $w(z)=p/\rho$ and other quantities.


The expected performance of SKA HI galaxy surveys in constraining dark energy was previously investigated by \citet{Abdalla:2009wr}. In this paper we update those results, using improved modelling of the number density and bias of the HI galaxy distribution, as well as more recent specifications for the various SKA configurations \citep{aaska14}. We provide the expected number counts and bias as a function of redshift and raw flux sensitivity, and map these on to specific SKA configurations. We then present Fisher forecasts for the BAO for each configuration, and use these to compare with the performance of other galaxy surveys.

\begin{table*}
\begin{center}
{\renewcommand{\arraystretch}{1.5}
\begin{tabular}{llccccrcccc}
\hline
\multirow{2}{*}{Telescope} & \multicolumn{1}{c}{Band~~~~} & {Target freq.$^{(a)}$} & {$T_\mathrm{inst}$$^{(b)}$} & {$N_{\rm dish}$$^{(c)}$} & $D_\mathrm{dish}$$^{(d)}$& $A_{\rm e}$$^{(e)}$ & {$N_b$$^{(f)}$} & Beam$^{(g)}$ & S$_{\rm rms}$$^{(h)}$\\
 & \multicolumn{1}{c}{[MHz]~~~~} & [MHz] & [K] & & [m] & [m$^2$] & & [deg$^2$] & [$\mu$Jy] \\
\hline
SKA1-MID & 950 -- 1670 & 1355 & 20 & 190 & 15 & 26,189 & 1 & 0.48 & 247\\
MID+MeerKAT & 950 -- 1670 & 1310 & 20 & 254 & -- & 32,144 & 1 & 0.51 & 202\\
SKA1-SUR  & 650 -- 1670$^{(i)}$ &  1300  & 30 & 60 & 15 & 8,482 & 36 & 18 & 1151\\
SUR+ASKAP & 650 -- 1670 & 1300 & 38/30$^{(j)}$ & 96 & -- & 11,740 & 36 & 18 & 1050/830$^{(j)}$\\
\hline
SKA2$^{(l)}$ & 480 -- 1290 & -- & -- & -- & -- & -- & -- & 30 & 16\\  
\hline
\end{tabular} }
\end{center}
\caption[x]{Specifications for the various SKA telescope configurations (see text and notes below for explanation). Values quoted at the middle of the band in order to compare directly to the SKA baseline design document \citep{dewdney2013ska1}.\\
{\bf Notes:} {\bf (a)} This is the frequency at which frequency-dependent quantities are calculated. For PAFs this is taken to be the critical frequency (below which the beam is constant).
{\bf (b)} Instrument temperature.
{\bf (c)} Number of dishes.
{\bf (d)} Dish diameter.
{\bf (e)} Total effective collecting area of the interferometer.
{\bf (f)} Total number of beams (feeds). 
{\bf (g)} Total FOV (primary beam, for one pointing), calculated at the target frequency. For combined telescopes, the smaller beam of the two is used. For PAFs, it is multiplied by the number of beams above the critical frequency.
{\bf (h)}  Flux rms for a frequency interval of 10\,kHz and 1\,hour integration using Eq. \eqref{rms}.
{\bf (i)} Only 500 MHz instantaneous bandwidth assumed.
{\bf (j)} The first value takes a weighted average of the SUR+ASKAP temperature, while the second assumes that the ASKAP PAFs are replaced to match SUR system temperature.
{\bf (l)} Values here are only indicative (see text).}
\label{tab:telescopes}
\end{table*}

\section{Telescope and Survey specifications}

In this section, we analyse the specifications and expected flux sensitivities of surveys with various SKA configurations.

The SKA will be built in two phases. Phase 1 will consist of three separate sub-arrays: SKA1-MID, SKA1-SUR and SKA1-LOW \citep{dewdney2013ska1}. MID and SUR are dish arrays equipped with the mid-frequency receivers ($\nu \lesssim 1.4$ GHz) necessary to detect HI emission at low/intermediate redshift, while LOW is an aperture array optimised for lower frequencies ($< 350$ MHz) and thus higher redshifts. We will concentrate on MID and SUR here, and their corresponding `precursor' arrays, MeerKAT and ASKAP, which they will be co-sited with, and which can be connected into the final Phase 1 systems. LOW will be capable of detecting HI emission only for $z \ge 3$, which will presumably be done most efficiently using intensity mapping rather than a galaxy survey, so we will not consider it here (although see e.g. \citealt{Villaescusa-Navarro:2014rra}).

The specifications of Phase 2 are less well-defined. While its target sensitivity has been given -- around $10\times$ that of MID or SUR at mid-frequency -- the receiver technology, field of view, and baseline distribution are not yet decided. As such, we can only speculate on these details here. To ``future-proof'' our results to some extent, in later sections we will present results for the HI galaxy number counts and bias as a function of raw flux sensitivity, as well as for individual experimental configurations. The former can easily be rescaled for the actual specifications of Phase 2 when they are announced, as well as for any other future radio experiment that targets HI.

\subsection{Flux sensitivity} \label{sec:flux-sens}

We begin by reviewing the basic flux sensitivity equation. The rms (root mean square) noise associated with the flux measured by an interferometer is
\begin{equation}
S_{\rm rms} \approx \frac{2 k_B T_{\rm sys}}{A_{e} \sqrt{2\delta\nu\, t_p}},
\end{equation}
for a telescope with system temperature $T_\mathrm{sys}$, total effective collecting area $A_{e}$, frequency resolution $\delta\nu$, and observation time per pointing $t_p$ ($k_B$ is the Boltzmann constant). We have assumed that the noise is Gaussian.
The extra factor of $1/\sqrt{2}$ comes from assuming a dual-polarisation receiver system. For a dish reflector, the effective collecting area is typically about 70\% of its total geometrical area.

The expression above gives the flux sensitivity for the telescope psf (point spread function); that is, the noise rms for an ``angular'' pixel set by the resolution of the interferometer (not to be confused with its field of view or primary beam). This calculation corresponds to the so called ``natural array'' sensitivity. If angular resolution is an issue, then some uniform weighting plus Gaussian tapering of the visibilities might be required in order to improve the psf. In that case, sensitivities will be reduced with respect to the target values quoted. 
This is mostly an issue for continuum surveys, where angular resolution is crucial for separating the galaxies, but not as much for a HI survey since galaxies can in principle be detected along the frequency direction too by resolving their HI line. Nevertheless, we will consider a range of values when analysing the cosmological performance to allow for differences in the final line-processed sensitivity.

The total system temperature is given by $T_{\rm sys}=T_{\rm inst}+T_{\rm sky}$,
where the contribution from the sky is $T_{\rm sky}\approx 60 \left(300\, {\rm MHz}/\nu\right)^{2.55}$ K, and
$T_{\rm inst}$ is the instrument temperature (which is usually higher than the sky
temperature for $\nu \gg 300$ MHz). For typical instrumental specifications, the
noise rms for the array can be written as
\be \label{rms}
S_{\rm rms} = 260\, {\rm \mu Jy}\left(\frac{T_{\rm sys}}{20\, \rm K}\right)
 \left(\frac{25,000\, \rm m^2}{A_e}\right)\times \left(\frac{10 \,{\rm kHz}}{\delta\nu}\cdot \frac{1 {\rm hr}}{t_{\rm p}} \right)^{1/2}. \nonumber
\ee
We will assume that the interferometer, {\it in a single pointing}, can observe the following sky area, corresponding to the primary beam or field of view of a dish:
\begin{equation}
\theta_{\rm B}^2 \approx \frac{\pi}{8} \left (\frac{1.3 \lambda}{D} \right )^2\ [{\rm sr}],
\end{equation}
where any efficiency factor has already been taken into account. This is valid for dishes with single feeds (single pixels) like MeerKAT and SKA1-MID. The ASKAP and SKA1-SUR dishes are equipped with Phased Array Feeds (PAFs), however, for which the situation is slightly more complicated. PAF systems are able to observe a total of $N_b$ beams, depending on the number of feeds, so that the total field of view should be $N_b\times \theta_{\rm B}^2$. While $\theta_{\rm B}^2$ increases with wavelength, the total effective PAF beam will remain constant above a certain critical wavelength, corresponding to where the individual sub-beams begin to overlap with one another.

The specifications for each SKA configuration are summarised in Table~\ref{tab:telescopes}, along with the expected flux rms for a one hour integration in a single pointing with a frequency resolution of 10 kHz. For SKA1-MID/SUR and their combination with MeerKAT/ASKAP, only Band 2 is considered, as the lower-frequency Band 1 will provide insufficient sensitivity for a HI galaxy survey. For the combined telescopes, only the overlapping band is given. Note however that the SKA1 baseline specifications suggest that the ASKAP PAFs should be replaced to match the SKA1-SUR band and instrumental temperature (taken to be 30 K). 

For SKA2, as mentioned above, we just assume 10 times the sensitivity of the Phase 1 configurations, leaving other aspects of the specification (e.g. system temperature, number of dishes) undefined. We must still choose a field of view (FOV) and bandwidth, however; reasonable estimates are a FOV about 20 times that of the Phase 1 configurations, and a bandwidth sufficient to cover $0.1 \le z \le 2.0$ (i.e. $1290 \ge \nu \ge 480$ MHz). The significantly larger FOV can be supported by various proposed technologies for Phase 2, e.g. MFAA,\footnote{\url{https://www.skatelescope.org/mfaa/}} while the Phase 1 dish arrays will already possess the technology required to cover the specified frequency range (albeit with lower sensitivity, limiting the useful minimum frequency for HI galaxy surveys).

\subsection{Survey specifications}
\label{survey}

To maximise its effectiveness, a balance must be found between the sensitivity of a survey and its area. In principle, wide surveys can probe larger volumes and thus sample a greater number of Fourier modes, but this comes at the cost of reducing sensitivity per pointing (for a fixed total survey time), thus increasing shot noise and reducing the maximum redshift that can be reached.

\begin{figure}
\vspace{-1em}
\includegraphics[width=0.5\textwidth]{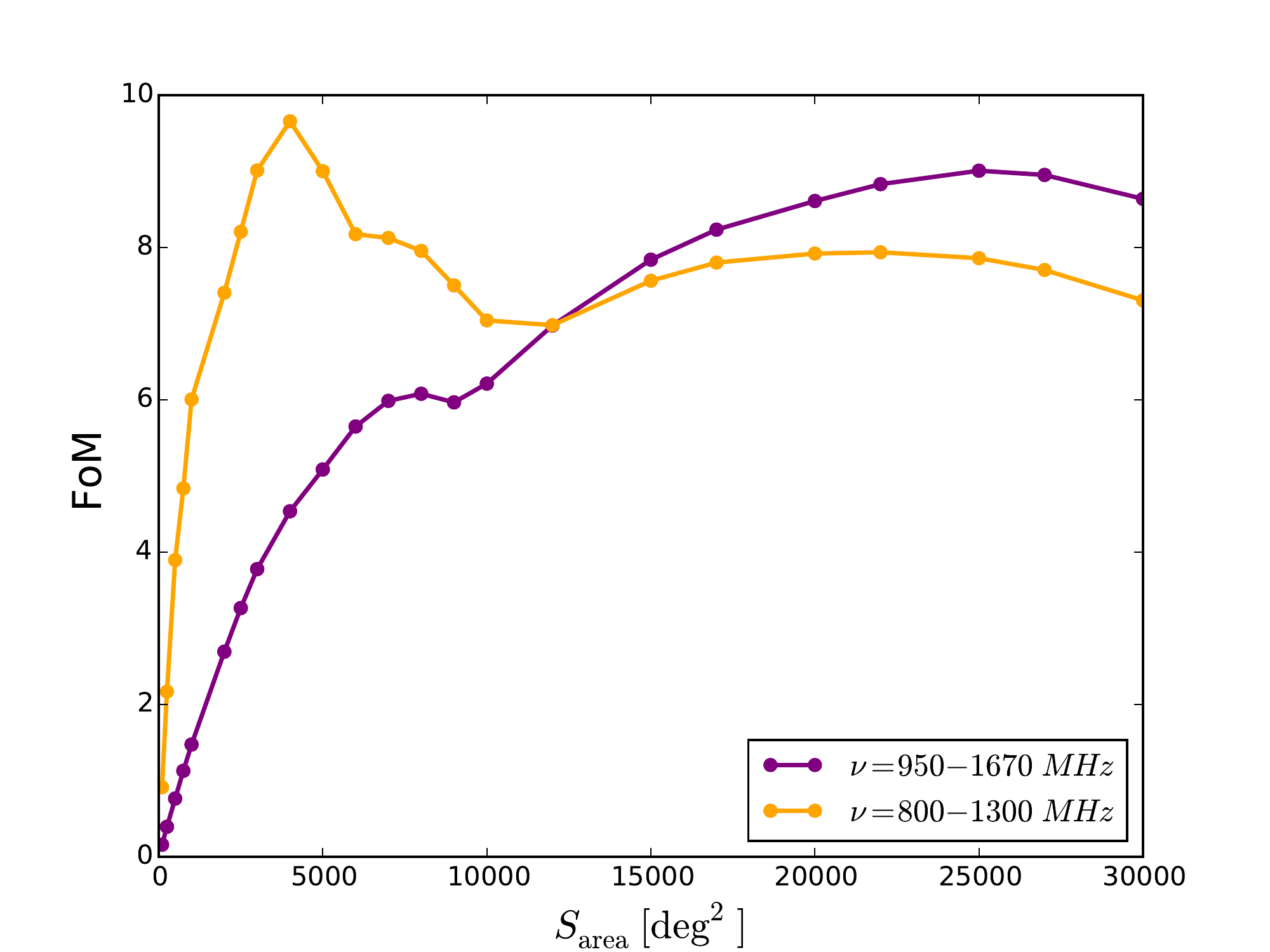}
\caption{Dark energy figure of merit versus survey area for SKA1, with different frequency ranges. A Planck CMB prior has been included in the FOM calculation, but the optimal survey areas are the same even if this is not included.}\vspace{-2em}
\label{fig:area_vs_fom}
\end{figure}

For a 10,000 hour survey, and the sensitivities given in Table \ref{tab:telescopes}, SKA2 will essentially be able to achieve a sample variance-limited ``full sky'' survey, so we set its survey area to 30,000 deg$^2$. For SKA1, however, the situation is less obvious. Using the Fisher matrix analysis described in the following sections, we searched for the optimal SKA1 survey area for our target science -- in this case, whatever maximises the dark energy figure of merit (FOM). We also considered two possible frequency intervals: the current SKA1 MID specification (950-1670 MHz) and a slightly ``deeper'' band (800-1300 MHz) with a maximum redshift of $\sim 0.8$. The results of the optimisation procedure are shown in Fig.~\ref{fig:area_vs_fom}.

For the lower frequency range, Fig.~\ref{fig:area_vs_fom} shows that the FOM is maximised for a survey area of around 5,000 deg$^2$. This where a balance between depth and width is reached -- the information gain from detecting the BAO at higher redshifts is traded-off against the larger sample variance due to the smaller area. Conversely, the higher frequency band is restricted to lower redshifts, limiting the maximum depth that can be achieved. This leads to a preference for larger areas ($\sim 25,000$ deg$^2$), although note that this survey would not be sample variance-limited as in the SKA2 case.

These results are specific to the figure of merit that we are optimising for. If we instead required a strong detection of the BAO at the highest redshift for the 950-1670 MHz band, for example, the optimal area would again be around 5,000 deg$^2$. Other issues can also be considered. For instance, we might want to ``piggy-back'' the HI galaxy survey on top of other surveys to optimise the overall observing time, which could drive us to use the 10,000 hours over $\sim 25,000$ deg$^2$. A ``full sky'' survey would also have the advantage of probing wide-angle effects.

In this work, we have opted to assume a 5,000 deg$^2$ area for SKA1 using the 800-1300 MHz band. Although the current specifications for, say, SKA1-MID Band 2, specify a minimum frequency of 950 MHz, the numbers are still under review, so that it is acceptable to assume that such a change could happen. Using a smaller 5,000 deg$^2$ area for SKA1 allows the survey to be sample variance-limited in every redshift bin, which also brings advantages in terms of dealing with possible systematics (e.g. it will be easier to deal with a 5-sigma detection threshold). The final specifications that we assumed are summarised in Table~\ref{tab:surveys}. 
\begin{table*}
\begin{center}
{\renewcommand{\arraystretch}{1.5} \begin{tabular}{cccccccc}
\hline
 Telescope & Redshift & Target freq. & Beam $[\mathrm{deg}^2]$ & $S_{\rm area}$ $[\mathrm{deg}^2]$ & $t_{\rm p}$ [hours] & S$^{\rm ref}_{\rm rms}$ [$\mu$Jy]\\
\hline
MID+MeerKAT$^{(a)}$ & 0.0 -- 0.78 & 1.0 GHz & 0.88 & 5,000 & 1.76 & 152\\
SUR+ASKAP$^{(b)}$ & 0.0 -- 1.19 & 1.3 GHz & 18 & 5,000 & 36 & 175/140$^{(c)}$\\
\hline
SKA2$^{(d)}$ & 0.1 -- 2.0~ & 1.0 GHz & 30 & 30,000 & 10 & 5.14\\
\hline
\end{tabular} }
\end{center}
\caption[x]{Survey specifications. We assume a total observation time of 10,000 hours. For MID+MeerKAT, a modification of band 2 is assumed (800-1300 MHz) in order to achieve the target redshift range.
Flux rms is calculated for a frequency interval of 10\,kHz. 
Values for the beam and flux sensitivity are quoted at the target frequency. \\
{\bf Notes:} {\bf (a)} Beam and time per pointing ($t_p$) are assumed to change as $\left(1\, {\rm GHz}/\nu\right)^2$ across the band, and the flux rms is assumed to change as $\nu/(1\, {\rm GHz})$.
{\bf (b)} Values calculated at the PAF critical frequency. Below that frequency, the values are assumed constant. Above it, the beam and $t_p$ are assumed to go as $1/\nu^2$, and the flux rms as $\nu$.
{\bf (c)} The first value takes a weighted average of the SUR+ASKAP temperature while the second value assumes that the ASKAP PAFs are replaced to match SUR system temperature.
{\bf (d)} Indicative; the beam and flux rms are assumed constant across the band.}
\label{tab:surveys}
\end{table*}

For a given survey area, $S_{\rm area}$ we will need approximately $S_{\rm area}/(\theta_{\rm B})^2$ pointings. The time per pointing $t_p$ is then related to the total integration time $t_{\rm tot}$ through
\begin{equation}
t_p=t_{\rm tot} \frac{(\theta_{\rm B})^2}{S_{\rm area}}.
\end{equation}
Since $(\theta_{\rm B})^2$ goes as $1/\nu^2$, this will increase the available integration time per pointing at lower frequencies; the flux rms is therefore proportional to the frequency, and so decreases for lower frequencies. The flux rms will remain constant below the critical frequency for PAFs, however, as explained above.
In order to cover the required survey area, we assume that the mosaicking (how we pack the pointings/beams) is done at the highest frequency used for the HI survey; that is, the telescope pointings are packed side by side at the highest frequency. This ensures that the full survey area is covered at the highest frequency, but means that the beams will overlap at lower frequency, reducing the survey efficiency.

\section{HI galaxy simulations}

Crucial ingredients to any cosmological calculation using galaxy surveys are the galaxy number density as a function of redshift and detection threshold, and the corresponding bias with respect to the underlying dark matter distribution. Analytical calculations, though possible, would have to rely on some relation between the HI luminosity for a given galaxy and its host dark matter halo. As such, they might fail to emulate the actual distribution unless properly calibrated to full simulations, as the HI luminosity can depend on other factors besides the halo mass.
\begin{figure}
\begin{center}
\includegraphics[width=0.5\textwidth]{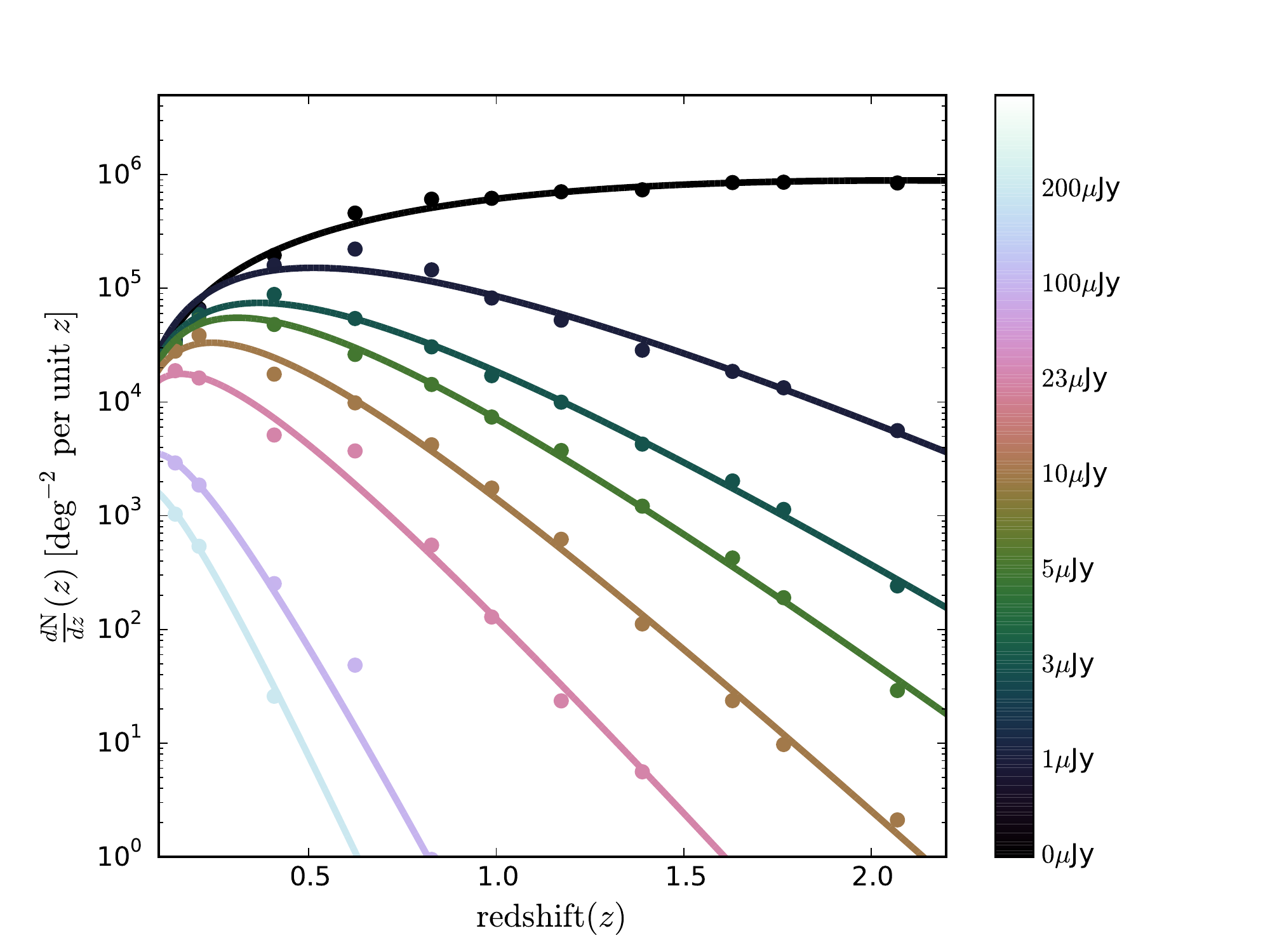}
\caption{HI galaxy redshift distribution, $dN/dz$, calculated from simulations (solid circles) and the corresponding fitting function, Eq.~\eqref{equ:dndz}. From top to bottom, the curves shown correspond to flux sensitivities $S_{\rm{rms}} = (0, 1, 3, 5, 10, 23, 100, 200) \ \mu{\rm Jy} $ (colour-coded according to the panel on the right).}
\label{fig:dNOverdz_fit_sax3}
\end{center}\vspace{-3em}
\end{figure}

Instead, to calculate the HI galaxy number density and bias as a function of the survey rms sensitivity $S_{\rm{rms}}$ we have used the S$^3$-SAX simulation.\footnote{\url{http://s-cubed.physics.ox.ac.uk/s3_sax}}
This simulation consists of a galaxy catalogue containing the position and
several astrophysical properties for objects in a mock observing cone. It was
produced by \citet{2009ApJ...703.1890O} by adding HI and CO properties to the galaxies
obtained by \citet{2007MNRAS.375....2D} through the post-processing of the Millennium
dark matter simulation \citep{2005Natur.435..629S}. Since each galaxy in the simulation has associated with it a HI luminosity and line profile, as well as a redshift, we can proceed to calculate the number of galaxies that one could expect to detect with a given survey.

\begin{figure}
\includegraphics[width=0.5\textwidth]{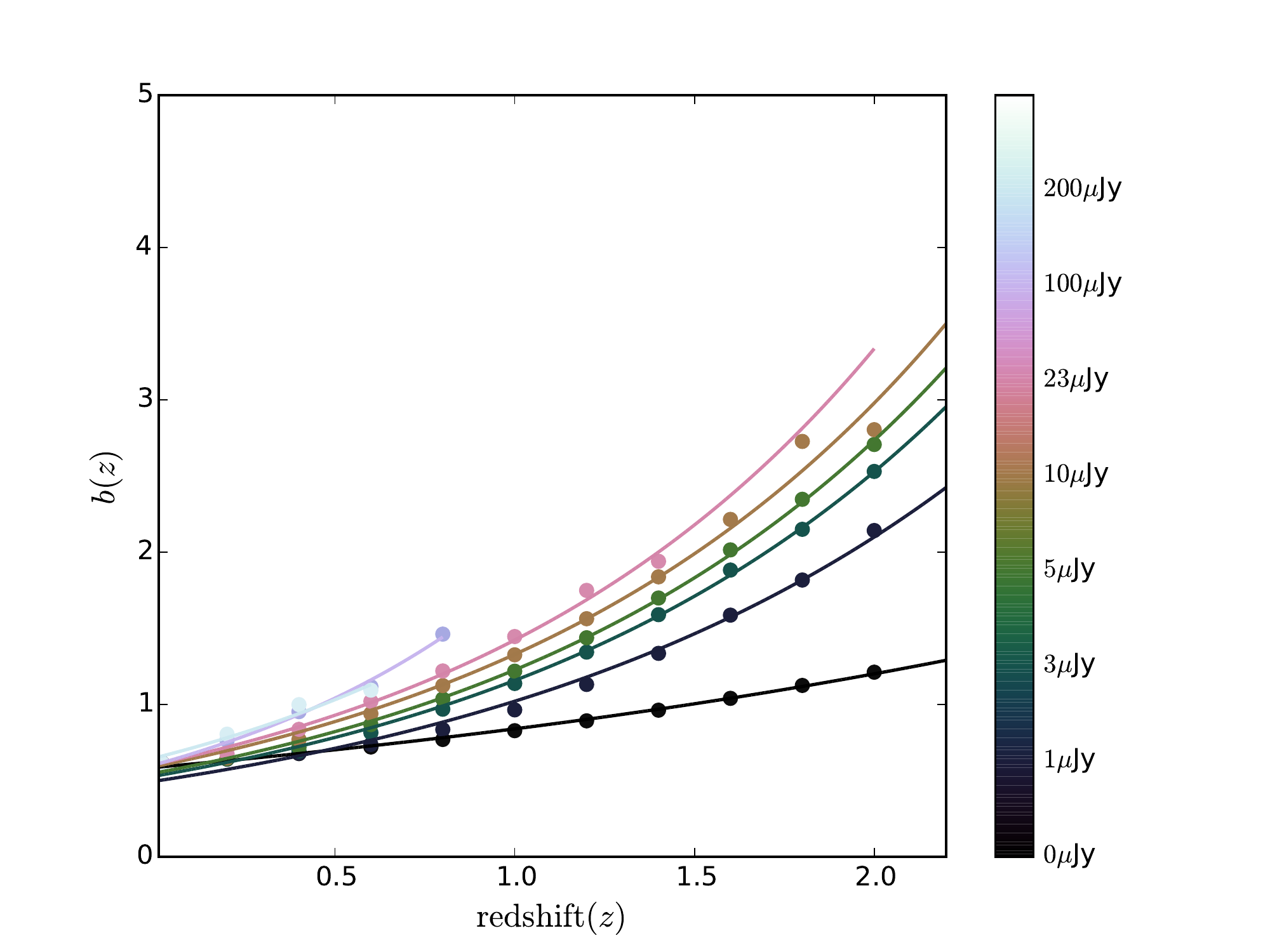}
\caption{HI galaxy bias for different $S_{\rm{rms}}$. Note that bias values for high flux rms are uncertain.
This has little impact, however, as shot noise will dominate at these sensitivities.} 
\label{fig:bias_fit_sax3}
\end{figure}

\subsection{HI galaxy number densities}

Detection of a HI galaxy relies on the measurement of its corresponding HI line profile. This is usually set by the galaxy rotation curve and the inclination angle at which the galaxy is observed. The largest line width will be obtained if we observe the spiral galaxy edge-on and the smallest when it is observed face-on. The choice of detection algorithm is crucial to the success of any large HI galaxy survey campaign, as it will determine the total number of galaxies detected and how clean that detection is, i.e. how well spurious detections (due to RFI, for instance), can be rejected. As such, the expected galaxy number density for a given survey is not simply a function of the flux sensitivity.

In this paper, we take the simple approach that at least two points on the HI line are required to be
measured in order for a galaxy to be detected. That is, the width of the line has to be larger than twice the assumed frequency resolution of the survey. The idea is to obtain information on the
typical line double peak (double horn) expected from HI galaxies due to their rotation. This will remove any galaxy that is seen face-on since it would just show as a
narrow peak, which could be confused with RFI. Typical line profiles have widths of tens of kilometres per second, which is fine for the radio telescopes we are considering, as resolutions of 10 kHz are easily achievable (corresponding to $\sim2$ km/s in the rest frame). 

Using the S$^3$-SAX database, we applied the following ``detection'' pipeline:
\begin{enumerate}
\item Take $z_{\rm A}$ (the apparent redshift, including Doppler correction) from the database.
\item Set the spectral resolution to $\delta V = 2.1(1+z_{\rm A})$ km/s, corresponding to a frequency resolution of $10\,$kHz (which was assumed for the sensitivity calculations).
\item
Take $w_{\rm P}\,$ (the line width between the two horns of the HI line profile, corrected for galaxy inclination) from the database, and select only galaxies with $w_{\rm P} > 2 \delta V$.
\item
Take $v_{\rm HI}\,$ (the velocity-integrated line flux of the HI line) from the database and select only galaxies where the flux$\, =  v_{\rm HI} /w_{\rm P}>N_{\rm cut} \times S_{\rm rms}/\sqrt{(w_{\rm P}/\delta V)}$. This corresponds to a detection threshold of $N_{\rm cut} \times 1 \sigma$ for the HI line.
\end{enumerate}
Note that $S_{\rm rms}$ is only the flux sensitivity -- the survey flux cut will be a factor of several above that (usually five or ten, depending on the chosen threshold), although the actual value is not straightforward to specify since it depends on the detection algorithm.

In order to be as general as possible, we give results for a range of $S_{\rm rms}$ values so that a simple interpolation can be used if there is a change in the survey specifications. 
We use the  formula of \citet{2009ApJ...703.1890O} to fit the $dN/dz$ data points from S$^3$-SAX:
\begin{equation}
\frac{{ d}N(z)/ {d}z}{{1\, \rm deg}^2}= 10^{c_1} z^{c_2} {\rm exp}\left( - c_3 z\right), 
\label{equ:dndz}
\end{equation}
where $c_i$ are free parameters. Note that $\frac{dN}{dz}$ is the number of galaxies per square degree and per redshift interval.
Figure~\ref{fig:dNOverdz_fit_sax3} shows the fitted curves and the simulated data points, and the fitted parameters are given in Table~\ref{table:free_parameters}.

\subsection{HI galaxy bias}\label{HI_galaxy_bias}

To calculate the galaxy bias using the SAX simulation, two approaches were considered. The most direct was to put the extracted HI galaxies in a box according to their redshift and position, and to then calculate the galaxy power spectrum. The bias squared is then the ratio of this power spectrum to the dark matter one at a given scale $k$. Ideally we would target large scales, to avoid non-linearities and shot noise contamination.
The initial box for the simulation was $500h^{-1}\,$Mpc, but this was further reduced along the line of sight to avoid cosmic evolution, which raises a problem for the bias extraction since linear modes with $k\lesssim 0.1 h/\,$Mpc will be affected by cosmic variance.

The other option was to calculate the HI galaxy bias using the dark matter halo bias. To that end, we need to extract from the simulation box, at a given redshift, the dark matter halo hosting each HI galaxy above the target flux cut. The HI bias can then be calculated using a weighted sum of the dark matter halo bias,
\be
b_\mathrm{HI}(z,S_{\rm rms}) \approx \sum_i b(z, M_i)\frac{N_i}{N_{\rm tot}},
\ee
where $b(z, M_i)$ is the halo bias for mass $M_i$ \citep{1999MNRAS.308..119S}, $N_i$ is the number of halos in the box with mass $M_i$ hosting HI galaxies above the detection threshold, and $N_{\rm tot}=\sum_i N_i$. This method is less affected by shot noise and does not suffer from the cosmic variance issues of the previous method. As such, in this paper we opted to calculate the bias following this second prescription.
The data points obtained from the simulation are shown in Fig.~\ref{fig:bias_fit_sax3} as a function of redshift for different $S_{\rm{rms}}$ sensitivities, and numerical values are given in Table A1 in the appendix. We fit the simulated data using
\be \label{bias}
b_\mathrm{HI}(z) = c_4 \exp({c_5z}),
\ee
and give the values of the best-fit parameters in Table~\ref{table:free_parameters}.

The galaxies used in the bias calculation are contained in small volumes between $\sim (60 / h)^3$ Mpc$^3$ (for $z\approx 0$) and $(175 / h)^3$ Mpc$^3$ (for $z \approx 2$) due to the size of the redshift bins considered. Given the much larger volumes probed by an experiment like the SKA, one would expect to find a number of halos larger than those contained in the simulation boxes. However, this should only have an impact for large flux cuts, which are dominated by shot noise anyway and so will have little consequence in terms of cosmological constraints.

For halos of a given mass, there is significant variation in the HI mass of the galaxies residing within them. This implies that some galaxies with considerably higher HI masses than the average will be found. The number of halos rapidly decreases with halo mass and redshift, however, and so the majority of galaxies with high HI masses will be found in modest halos with modest bias. The fraction $M_\mathrm{HI}/M_\mathrm{halo}$ has also been shown to rapidly decrease with increasing halo mass for halos with masses above $10^{12} M_\odot$ \citep{2014arXiv1409.1574P}, so even very massive halos are likely to have modest HI masses of the order of $10^9 M_\odot$ on average. This has the effect of introducing an effective upper limit to the bias at each redshift, which we estimated to be only slightly higher than the maximum values we were able to obtain from the simulation. As such, at each redshift one can assume that the bias remains constant for values of $S_{\rm rms}$ higher than the maximum that could be extracted from the simulation.

For HI masses below $10^9 M_\odot$, locally-measured HI luminosity functions seem to imply many more galaxies than predicted by the simulation, suggesting that low mass galaxies are more HI rich than previously thought \citep{2014arXiv1409.1574P}. If this is the case, the bias will be smaller than predicted here for small values of $S_{\rm rms}$ (e.g. $\lesssim 1$ $\mu$Jy). This result is subject to completeness uncertainty and cosmic variance, however, and is yet to be confirmed \citep{2013ApJ...766..137O}. Conversely, DLA observations (though model-dependent, and suffering from several uncertainties) are so far consistent with our predictions for the HI bias \citep{2012JCAP...11..059F}.

\begin{table}
\vspace{1.5em}\begin{centering}
{\renewcommand{\arraystretch}{1.5}
\begin{tabular}{rrrrrr}
\hline 
\multicolumn{1}{c}{$S_{\rm rms}$} & \multicolumn{1}{c}{c$_1$} & \multicolumn{1}{c}{c$_2$} & \multicolumn{1}{c}{c$_3$} & \multicolumn{1}{c}{c$_4$} & \multicolumn{1}{c}{c$_5$} \\ 
\hline 
0.0 & 6.21 & 1.72 & 0.79 & 0.5874 & 0.3577 \\
\hline
1.0  & 6.55 & 2.02 &  3.81 & 0.4968 &    0.7206 \\
\hline
3.0 & 6.53 & 1.93&  5.22 & 0.5302 &     0.7809 \\
\hline
5.0  & 6.55 & 1.93 & 6.22 & 0.5504 &    0.8015 \\
\hline
6.0 & 6.58 & 1.95 &  6.69 & 0.5466 &  0.8294 \\
\hline
7.3   & 6.55 & 1.92 &  7.08 & 0.5623 & 0.8233 \\
\hline
10 &  6.44 & 1.83 &  7.59 & 0.5928 &     0.8072 \\
\hline
23 & 6.02  & 1.43 &  9.03 & 0.6069 &  0.8521 \\
\hline
40 & 5.74 & 1.22 &  10.58 & 0.6280 &    0.8442 \\
\hline
70  &5.62 & 1.11 &  13.03 & 0.6094 &    0.9293 \\
\hline
100 &  5.63 & 1.41 & 15.49 & 0.6052 & 1.0859 \\
\hline
150  & 5.48 & 1.33 & 16.62 & 0.6365 & 0.9650 \\
\hline
200 & 5.00 &1.04 & 17.52 & \multicolumn{1}{c}{---} & \multicolumn{1}{c}{---} \\
\hline
\end{tabular} }
\caption{Best-fit parameters for the number density and bias fitting functions, Eqs. (\ref{equ:dndz}) and (\ref{bias}), for different flux limits. $S_{\rm{rms}}$ is measured in $\mu$Jy.}
\label{table:free_parameters}
\end{centering}
\end{table}


\section{Cosmological Performance}

In this section, we use Fisher forecasts to compare the ability of the proposed SKA HI galaxy surveys to constrain various cosmological quantities. Our focus is on the detection of the BAO feature, which we use as a figure of merit owing to its status as arguably the cleanest \citep{2010ApJ...720.1650S, 2011ApJ...734...94M} and most `standard' observable targeted by cosmological large-scale structure surveys. Constraints on the dark energy equation of state parameters, $w_0$ and $w_a$, are also presented. We take the Planck best-fit flat $\Lambda$CDM model \citep{Ade:2013zuv} as our fiducial cosmology, with 
$h=0.67$, $\Omega_{{\rm cdm}} = 0.267$, $\Omega_{\rm b}= 0.049$, $n_s=0.962$, and $\sigma_8=0.834$.
\vfill

\subsection{SKA assumed sensitivities}

Our forecasts follow the specifications given in Table \ref{tab:surveys}, with the sensitivities obtained for a total observation time of 10,000 hours, and a survey area of 5,000 deg$^2$ for SKA1 and 30,000 deg$^2$ for SKA2. For each configuration we also considered `optimistic' and `pessimistic' variations, which are intended to bracket the possible range of flux sensitivities once HI modelling uncertainties and possible changes to the instrumental design are taken into account.

\begin{figure}
\begin{center}
\vspace{-1em}
\includegraphics[width=0.5\textwidth]{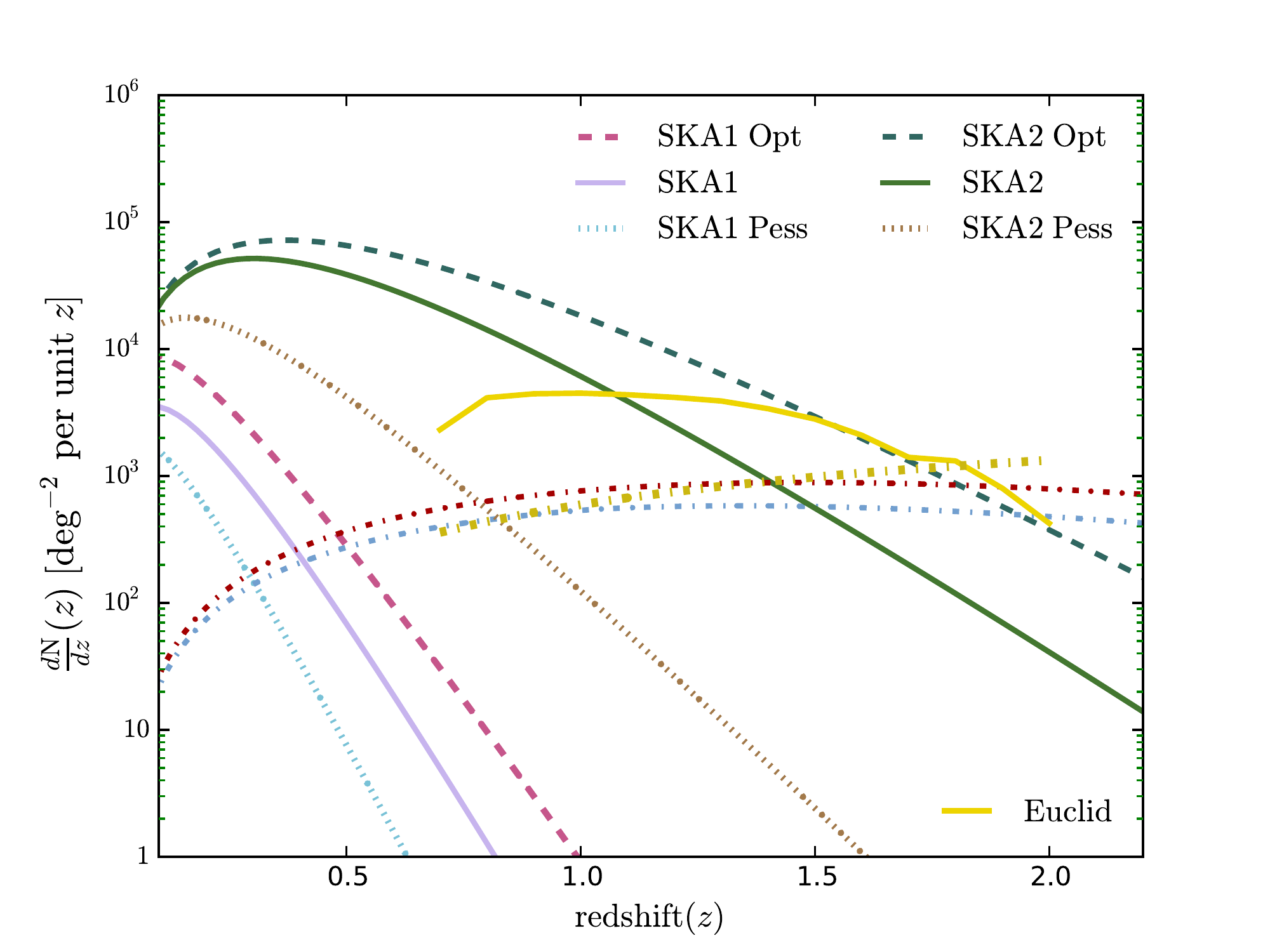}
\caption{Number densities for the optimistic, reference, and pessimistic cases of SKA1 and 2, compared with Euclid. Dashed-dotted lines show the number density at which $n(z) b^2(z) P(z, k_\mathrm{max}) = 1$ for the various surveys, with $k_\mathrm{max} \approx 0.2 \,h /\mathrm{Mpc}$. When $dN/dz$ is above this line, sample variance dominates the shot noise for all $k < k_\mathrm{max}$; the point at which it dips below the line is effectively the maximum redshift of the survey.}\vspace{-2em}
\label{fig:cosmic_variance}
\end{center}
\end{figure}

For SKA1, we take the flux rms at the target frequency of 1 GHz to be $S^{\rm ref}_{\rm rms} = 70/150/200\, \mu$Jy (opt./ref./pess.). The optimistic scenario is roughly equivalent to taking the reference flux for SKA1-MID+MeerKAT (152 $\mu$Jy), but assuming that the detection threshold would be set at the 5$\sigma$ level. For SKA2, in lieu of any other information about its design we take the flux rms to be constant across the band, with $S^{\rm ref}_{\rm rms} = 3.0/5.4/23\, \mu$Jy (opt./ref./pess.).

The frequency/redshift interval for SKA1 is taken to be compatible with SKA1-SUR + ASKAP Band 2 or a modification of SKA1-MID + MeerKAT Band 2 as explained in section \ref{survey}, e.g. 800-1300 MHz. We ignore Band 1, since above $z\sim 0.8$ one cannot detect enough galaxies for cosmological purposes with SKA1 sensitivities anyway. Note that both MID and SUR have similar sensitivities for the HI galaxy survey we are describing, although the current SUR band 2 definition is more optimal for this.
For SKA2, we take the $z$ range given in Table~\ref{tab:surveys}.

The number density and bias scale with frequency/redshift, as explained in Section \ref{sec:flux-sens}. We take this into account by interpolating between the best-fit sensitivity curves shown in Figs. \ref{fig:dNOverdz_fit_sax3} and \ref{fig:bias_fit_sax3}, as a function of redshift. The interpolation also allows us to factor in possible changes to the flux cut (galaxy detection threshold). For a given survey, the flux rms therefore scales as
\be
S_{\rm rms} = S^{\rm ref}_{\rm rms}\frac{N_{\rm cut}}{10} \frac{\nu_{21}}{\nu_c} (1 + z)^{-1},
\ee
where $\nu_{21}$ is the rest frame frequency of the $21$ cm line, $S^{\rm ref}_{\rm rms}$ is the reference flux sensitivity quoted in the tables, $N_{\rm cut}$ is the threshold above which galaxies are taken to be detected, in multiples of the noise rms, and $\nu_c$ is the target/critical frequency at which $S^{\rm ref}_{\rm rms}$ was calculated (1.0 GHz for MID and 1.3 GHz for SUR). Note that for SUR (PAFs), the flux $S_{\rm rms}$ will remain constant for frequencies below $\nu_c$.
 
As mentioned above, we assume that the reference experiment for SKA1 has non-PAF receivers (i.e. SKA1-MID + MeerKAT). For SKA2 we take the flux to be constant with redshift, also as discussed above. Then we correct for  number density and bias by interpolating Eqs.~\eqref{equ:dndz} and \eqref{bias} using the values in Table~\ref{table:free_parameters}.
The resulting best-fit parameters for the number density and bias functions are given in Table~\ref{tab:corrected_nz}. The redshift distribution for the target surveys is shown in Fig. \ref{fig:cosmic_variance}, and compared to the limit below which the survey becomes shot noise-dominated.

\begin{figure*}
\begin{center}
\includegraphics[height=7cm,width=8.5cm]{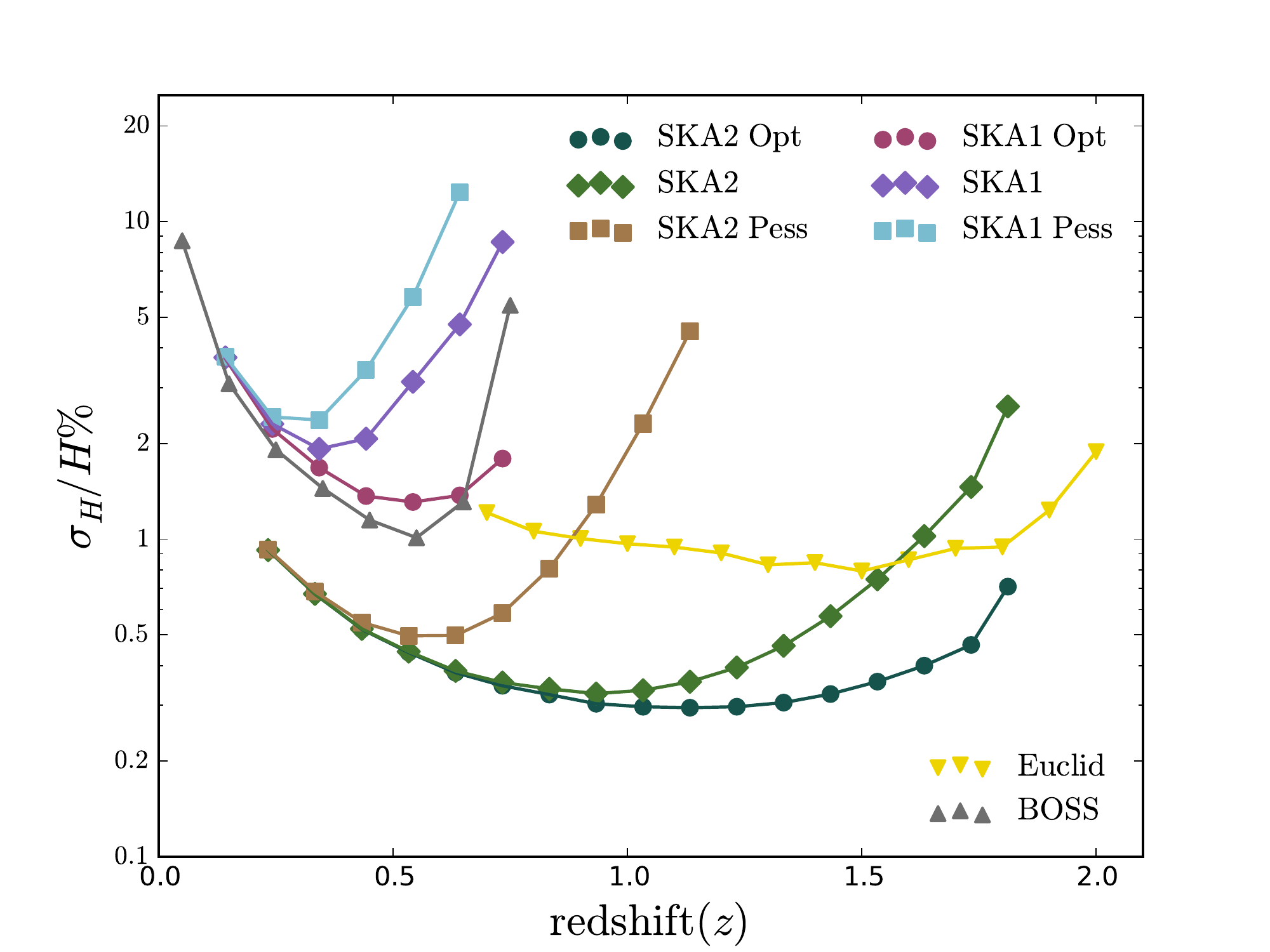}
\includegraphics[height=7cm,width=8.5cm]{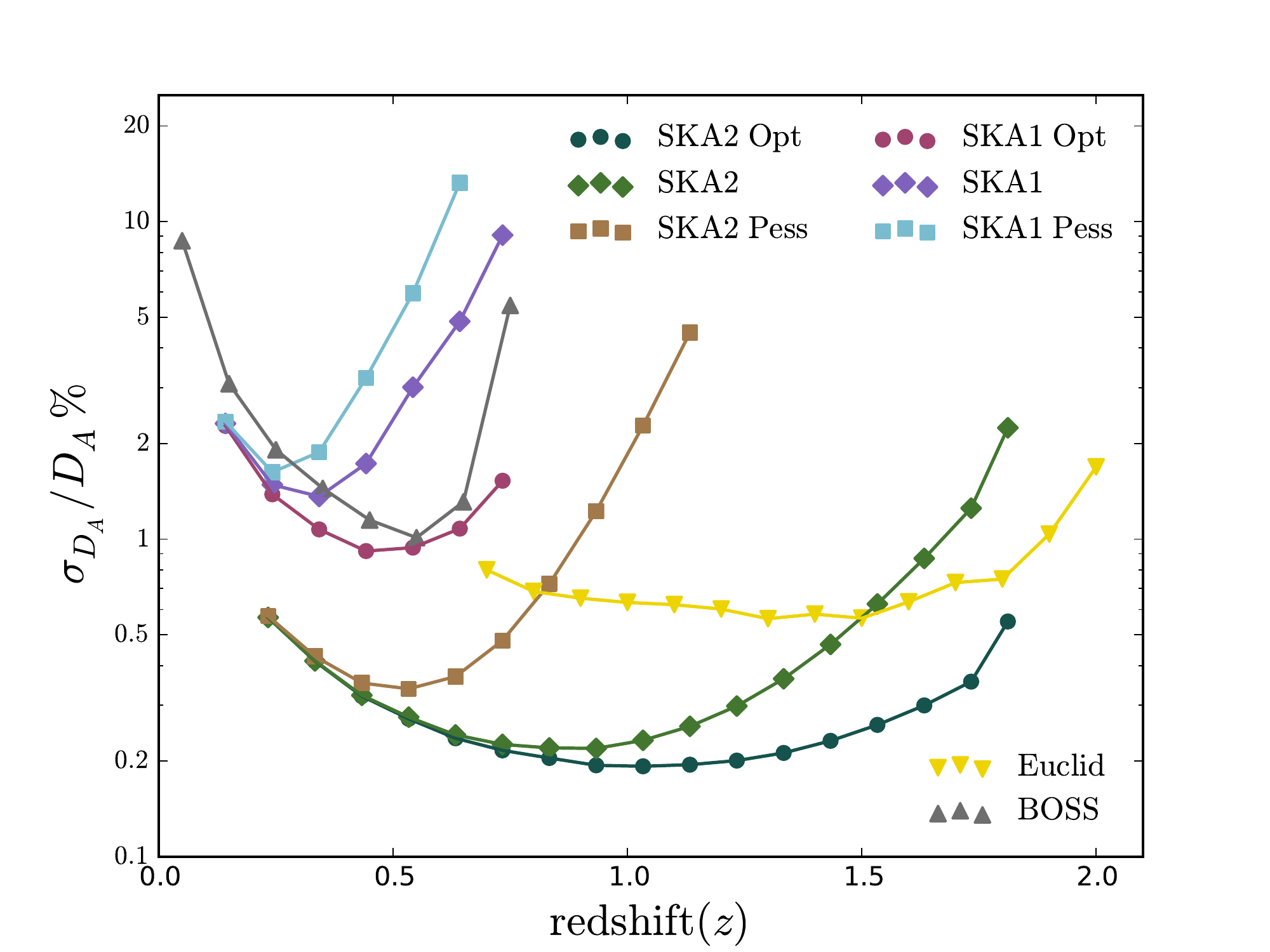}
\caption{Forecast fractional errors on the expansion rate, $H(z)$, and angular diameter distance, $D_A(z)$, from BAO measurements with the various surveys. The redshift binning is fixed at $\Delta z = 0.1$ for all experiments.}
\label{fig:fraction}
\end{center}
\end{figure*}

\begin{table}
\hspace{-1.5em}
{\renewcommand{\arraystretch}{1.3}
\begin{tabular}{llrrrrrr}
\hline 
 & & $S_{\rm rms}^{\rm ref}$ & \multicolumn{1}{c}{c$_1$} & \multicolumn{1}{c}{c$_2$} & \multicolumn{1}{c}{c$_3$} & \multicolumn{1}{c}{c$_4$} & \multicolumn{1}{c}{c$_5$} \\ 
\hline
\multirow{3}{*}{SKA1} & opt. & 70  & 5.253 & 0.901 & 7.536 & 0.628 & 0.819 \\
 & ref. & 150 &5.438 & 1.332 & 11.837 & 0.625 & 0.881 \\
 & pess. & 200 &5.385 & 1.278 & 14.409 & 0.646 & 0.896 \\
\hline
\multirow{3}{*}{SKA2} & opt. & 3.0 &6.532 & 1.932 & 5.224 & 0.530 & 0.781 \\
 & ref. & 5.4&6.555 & 1.932 & 6.378 & 0.549 & 0.812\\
 & pess. & 23.0& 6.020 & 1.430 & 9.028 & 0.607 & 0.852\\
\hline
\end{tabular} }
\caption{Fitted parameters for the galaxy number density and bias, for the frequency-corrected $S_{\rm rms}$ of the various experiments. The flux rms at the reference frequency, $S^{\rm ref}_{\rm rms}$, is in $\mu$Jy, while the fitted coefficients are dimensionless.}
\label{tab:corrected_nz}
\end{table}

\subsection{Fisher forecasts} \label{sec:fisher}

We use the Fisher forecasting technique to estimate how well the SKA surveys will be able to measure the BAO scale, and thus the various cosmological parameters. The first step is to construct the Fisher matrix, which is derived from a Gaussian approximation of the likelihood, evaluated for a set of fiducial parameters. For a spectroscopic galaxy redshift survey, the Fisher matrix in a single redshift bin is
\be
F_{ij} = \frac{1}{2} V_\mathrm{sur} \int \frac{d^3k}{(2\pi)^3} \frac{\partial \log C^T}{\partial \theta_i} \frac{\partial \log C^T}{\partial \theta_j},
\ee
where $\{\theta_i\}$ are the cosmological parameters of interest, $V_\mathrm{sur}$ is the comoving volume of the redshift bin, and we have neglected redshift evolution within the bin. The total variance of the measured fluctuations in the galaxy distribution is $C^T = P(\mathbf{k}, z) + 1/n(z)$, where $P(\mathbf{k}, z)$ is the redshift-space galaxy power spectrum, and $1/n(z)$ is the inverse of the galaxy number density, which acts as a shot noise term. Only the power spectrum depends on the cosmological parameters, so we can write
\bea
F_{ij} = \frac{V_{\rm sur}}{8 \pi^2}
\int_{-1}^{1}d\mu \int_{k_{\rm min}}^{{k_{\rm max}}} k^2 dk \left(\frac{nP}{1+nP}\right)^2 \frac{\partial \log P}{\partial \theta_i} \frac{\partial \log P}{\partial \theta_j}, \nonumber
\eea
where $\mu = \cos \theta$ is the cosine of the angle between the line of sight and the Fourier mode $\mathbf{k}$. We fix the lower integration limit to $k_{\rm min} = 10^{-3}h\,$Mpc$^{-1}$, and discard all information from modes beyond a non-linear cutoff scale,
\begin{equation}
k_{\rm max} = k_{\rm NL, 0} \left( 1+z \right)^{2/(2+n_s)},
\end{equation}
where $k_{\rm NL, 0} \simeq 0.2 h\,{\rm Mpc}^{-1}$ \citep{Smith:2002dz}.

We adopt a simplified `wiggles-only' approach to deriving BAO constraints, where only derivatives of the (Fourier-space) BAO feature are included in the Fisher matrix calculation. We first calculate the full (isotropic) power spectrum, $P(k, z)$, for the fiducial cosmology using CAMB \citep{Lewis:1999bs}, and then separate it into smooth and wiggles-only components such that \citep{Bull:2014rha}
\begin{equation}\label{fbao_of_k}
P(k, z) = \left [ 1 + f_\mathrm{BAO}(k) \right ] P_\mathrm{smooth}(k, z).
\end{equation}
If the actual cosmology differs from the fiducial cosmology, the observed wavenumber, $k$, of a feature in the isotropic power spectrum will be shifted according to \citep{Blake:2003rh}
\be
k = \sqrt{k^2_\perp ( D_A^\mathrm{(fid.)} / D_A )^2 + k^2_\parallel ( H / H^\mathrm{(fid.)} )^2}.
\ee

Since our aim is to provide a consistent comparison of the performance of various surveys, rather than to give high-precision forecasts on a large set of parameters, we make a number of simplifying assumptions: we ignore redshift-space distortions, non-linear effects, and uncertainty in both the bias and acoustic scale, and assume that the cosmological information encoded by the BAO feature comes entirely from the shift in $k$. We can then write
\be
\frac{\partial \log P}{\partial \theta} \approx \left [1 + f_\mathrm{BAO}(k) \right ]^{-1} \frac{d f_\mathrm{BAO}}{d k} \frac{d k}{d\theta}
\ee
where, following \citet{Seo:2007ns}, we work in terms of the parameters $\theta \in \{ \log D_A, \log H\}$, so that the Fisher integral factorises into a simple $2\times 2$ matrix of analytic angular integrals multiplied by the (scalar) $k$ integral. This calculation includes the cross-correlation between $D_A$ and $H$.  Because we are neglecting a number of nuisance parameters and other effects, our forecasts could be interpreted as somewhat optimistic -- although note that we are using only the information encoded in the BAO wiggles, which is quite insensitive to such effects (e.g. \cite{2012MNRAS.427.3435A}).

It is useful to project the constraints on $D_A$ and $H$ to various basic cosmological parameters. We first write the expansion rate and angular diameter distance as
\bea
H(z) &=& H_0 \left ( \Omega_{\rm m} (1+z)^{3} + \Omega_K(1+ z)^2 + \Omega_{\rm DE}(z) \right )^\frac{1}{2} \nonumber \\
D_{A}(z) &=& \frac{c}{H_0}\frac{(1+z)^{-1}}{\sqrt{-\Omega_K}}\sin\left(\sqrt{-\Omega_K} \int_0^z \frac{dz'}{E(z')}\right), \nonumber
\eea
where $\Omega_{\rm m}=\Omega_{{\rm cdm}} + \Omega_{\rm b}$ is the total matter density, $\Omega_K  = 1- \Omega_{\rm m} - \Omega_\mathrm{DE,0}$, is the spatial curvature, $E(z) = H(z)/H_0$ is the dimensionless expansion rate, and the dark energy evolution is given by
\begin{equation}\label{Fz}
\Omega_\mathrm{DE}(z)= \Omega_\mathrm{DE, 0} \,\exp \left [{\int_0^z \frac{3[1+w(z')]}{1+z'}dz'} \right ].
\end{equation}
We also define $H_0 = 100 h$ km/s/Mpc, and adopt the commonly used parametrisation of the dark energy equation of state, $w(z) \approx w_0 + w_a z/(1+z)$. The full set of parameters that we consider is then $\theta^\prime = \{ w_0, w_a,\Omega_{\rm cdm}, \Omega_{\rm b}, \Omega_K, h\}$, with the Fisher matrix found by projecting from the original $2 \times 2$ matrix and summing over redshift bins,
\begin{equation}\label{Eq:F_ab}
{F}^\prime_{\alpha\beta}=\sum_{ij,n} \left .{\frac{\partial \theta_i}{\partial \theta^\prime_\alpha}}{\frac{\partial \theta_j}{\partial \theta^\prime_\beta}}\right |_{z_n}\!\! F_{ij}(z_n).
\end{equation}
Finally, we add the Planck CMB prior Fisher matrix from \citet{2013LRR....16....6A} to represent the high-$z$ constraints that will be available.


\subsection{Comparison with previous results and future experiments}\label{sec:comparison}



\begin{figure}
\begin{center}
\includegraphics[height=8.0cm,width=9.0cm]{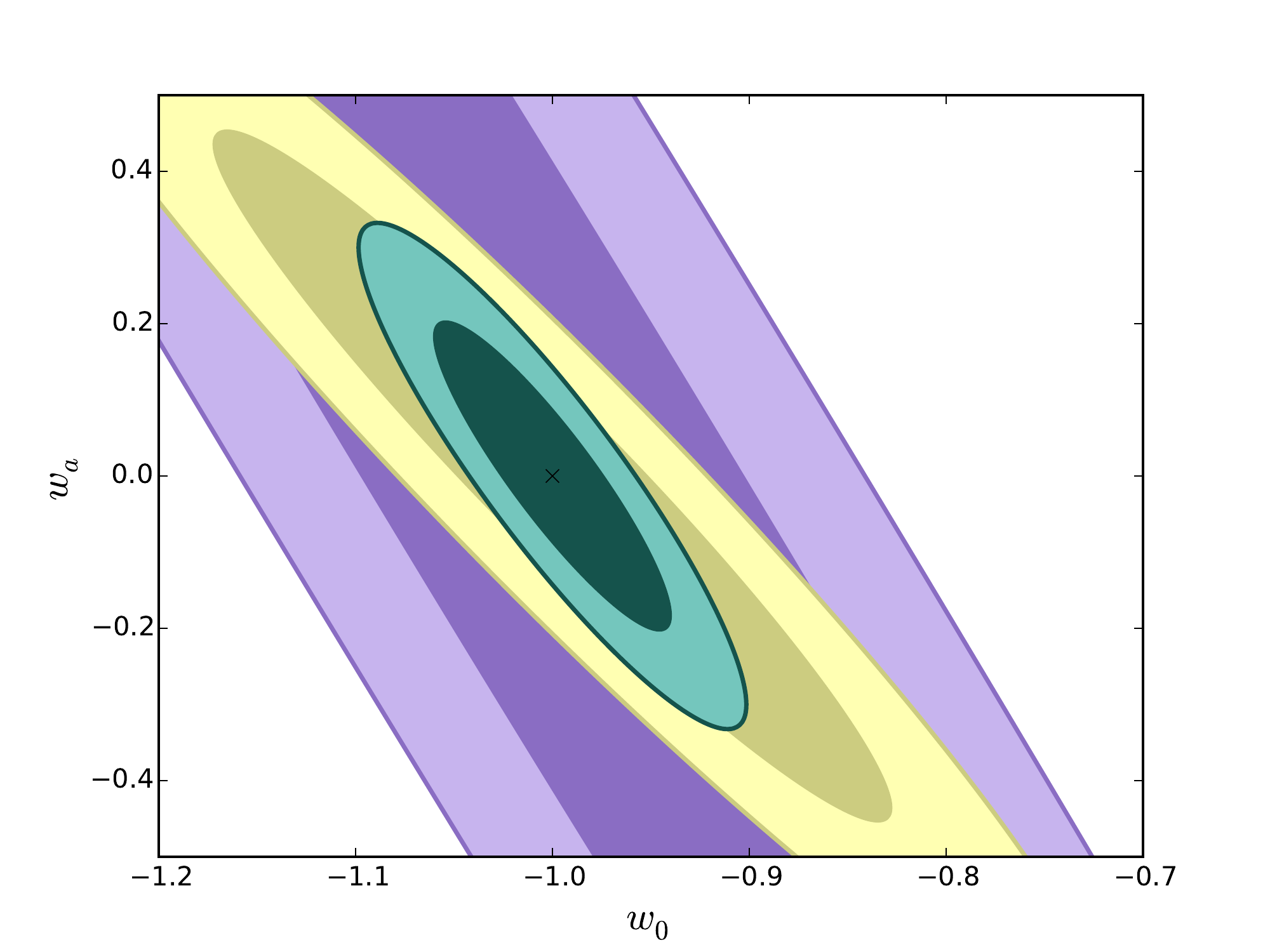} 
\caption{Forecast constraints ($1$ and $2\sigma$ contours) on the dark energy equation of state parameters, $w_0$ and $w_a$, for the reference cases of both SKA1 (purple, largest) and SKA2 (green, smallest), compared to a Euclid-like $H\alpha$ galaxy survey (yellow, intermediate). A Planck CMB prior has been included for all three experiments.}
\label{Fig:w_wa}
\end{center}
\end{figure}

\begin{figure}
\begin{center}
\includegraphics[height=8.0cm,width=9.5cm]{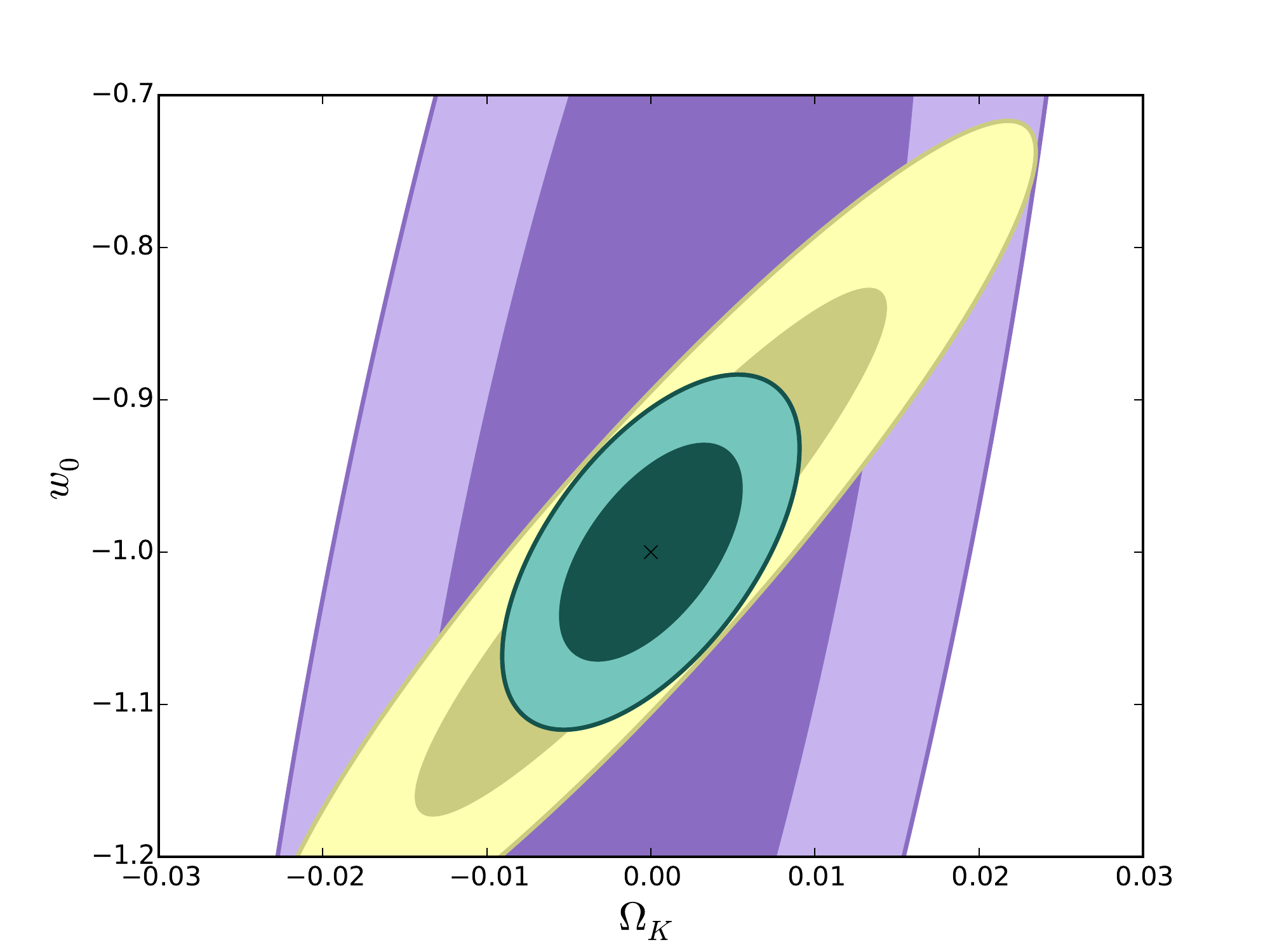} 
\caption{Forecast constraints ($1$ and $2\sigma$) on $w_0$ and $\Omega_K$ for the references cases of SKA1 (purple, largest) and SKA2 (green, smallest), compared with a Euclid-like H$\alpha$ galaxy survey (yellow, intermediate). A Planck CMB prior has been included for all experiments.}
\label{Fig:w_ok}
\end{center}\vspace{-1em}
\end{figure}

The results of our Fisher forecasts are shown in Figs.~\ref{fig:fraction} -- \ref{Fig:w_ok} and Table \ref{tab:marginals}. For comparison, we have also included forecasts for (a) a future optical/near-infrared H$\alpha$ galaxy survey with similar specifications to Euclid, using the number counts and bias model for the reference case described in \citet{2013LRR....16....6A}, and  (b) the BOSS LRG galaxy survey, using the specifications in \cite{2013AJ....145...10D}, with a total of 1.5 million galaxies out to $z \lesssim 0.75$, and with a bias of $b \approx 2$.


As can be seen from Fig. \ref{fig:fraction}, an SKA1 galaxy survey will offer -- at best -- only slight improvements over existing experiments at low redshift ($z \lesssim 0.7$). Indeed, from Table~\ref{tab:marginals} it can be seen that BOSS outperforms SKA1, although this is predominantly due to the larger assumed bias.

SKA1 should still significantly improve the cosmological constraints at low redshift, however, for the simple reason that it will cover a mostly independent survey area to existing experiments like BOSS and WiggleZ, thus increasing the total volume surveyed overall.

The picture is considerably more interesting for SKA2, which will be capable of performing a sample variance-limited survey over $\sfrac{3}{4}$ of the sky from $0.3 \lesssim z \lesssim 1.5$ in the reference case (increasing to $z \approx 2.0$ in the optimistic case). This will constitute the final word in spectroscopic redshift surveys in this redshift range, as there is little prospect of covering a greater survey area in the future. As shown in Fig. \ref{fig:fraction}, the SKA2 reference case is forecast to provide measurements of $H(z)$ and $D_A(z)$ to better than $0.5\%$ and $0.3\%$ precision respectively, out to $z \approx 1.3$. This significantly outperforms future H$\alpha$ surveys such as Euclid, which has half the survey area (and approximately double the errors) over the same range. This is contingent on performing at least as well as the reference case, however; the pessimistic case would only be competitive with Euclid out to $z \simeq 0.8$.

Even in the reference case, measurements above $z\sim 1.5$ would be difficult, as the HI source density falls too low (contrary to what has been forecast for Euclid, for instance). Note that the HI source density at $z > 1$ flattens as $S_{\rm rms} \to 0$ however (Fig.~\ref{fig:dNOverdz_fit_sax3}), suggesting that a sufficiently deep HI survey could produce precision constraints out to substantially higher redshift, at least in principle.


Figs.~\ref{Fig:w_wa} and \ref{Fig:w_ok} show forecasts for the equation of state and spatial curvature parameters for the reference cases of the various surveys. These were derived by projecting the $(H, D_A)$ Fisher matrices, including the cross-correlation terms, to the parameter set described in Sect. \ref{sec:fisher}, and then adding a Planck CMB Fisher matrix prior. Corresponding marginal errors are given in Table~\ref{tab:marginals} for the same parameters, for all cases. As before, SKA2 outperforms Euclid by a factor of around 2, reflecting its having double the survey area, as well as a further improvement due to  its 4 additional redshift bins below Euclid's minimum redshift. In terms of the dark energy figure of merit, defined as \citep{BASSETT2011, Coe:2009xf}
\be
{\rm FOM} = 1 / \sqrt{{\rm det}( \left . F^{-1} \right |_{w_0, w_a} )}
\ee
(equivalent to the inverse of the area of the $1\sigma$ $(w_0, w_a)$ ellipse), the SKA2 reference case performs around $4\times$ better than Euclid, and some $60\times$ better than SKA1 (opt. case).

Note that our forecasts are only intended for comparison of the various surveys. In reality, systematic effects (radio interference, the efficiency of source extraction algorithms, contamination by foreground emission, non-linearities, modelling errors etc.) should further affect the survey performance. We have concentrated exclusively on the BAO wiggles in our forecasts, however, which are hoped to give constraints more insensitive to such systematics. On the other hand, other observables (e.g. redshift space distortions) can also be measured, significantly improving the constraints on some parameters.

Leaving these issues aside, our calculations predict that the SKA2 (reference case) survey will be sample variance-limited over a significant fraction of the redshift range that is important for dark energy (i.e. $z \lesssim$ 2). As a result, it can come remarkably close to what would be possible with a `perfect' noise-free HI survey over the same area (represented by the $S_{\rm rms}=0$ entry in Table~\ref{tab:marginals}); the $1\sigma$ errors on $w_0$ and $w_a$ are only $\sim 1.5\times$ larger than their `noise-free' values, for example, and even in the pessimistic case they are still only $\sim 3\times$ larger.



\citet{Abdalla:2009wr} also investigated how well the SKA can measure the BAO scale and dark energy parameters. Our work differs from theirs in various aspects.  They used an analytical HI evolution model relying on prior knowledge of the star formation rate (SFR) and overall mass density of neutral hydrogen at a specific redshift, functions which depend on fitting formulas. We use a more realistic simulation to estimate the number counts, which we consider to be an improvement as our simulation relies on more physical properties, making our predictions more reliable.  The difference between the two sets of results can be seen by comparing the number counts (Fig.~\ref{fig:dNOverdz_fit_sax3}). For example, while they have a sharp curve as a function of redshift, ours decreases more gradually.\footnote{See Fig.~3 of \citet{Abdalla:2009wr}.} The second important difference is that while they assumed $b=1$, the bias in our simulation was a function of redshift, and was dependant on the frequency-corrected $S_{\rm rms}$ value (see Fig.~\ref{fig:bias_fit_sax3}).

\begin{table*}
{\renewcommand{\arraystretch}{1.5}
\begin{tabular}{ lrllllllr}
\hline 
 & \multicolumn{1}{c}{$S_{\rm rms}$ [$\mu$Jy]} & \multicolumn{1}{c}{$\sigma_{w_0}$} &  \multicolumn{1}{c}{$\sigma_{w_a}$} &  \multicolumn{1}{c}{$\sigma_{\Omega_{\rm cdm}}$} & \multicolumn{1}{c}{$\sigma_{\Omega_{\rm b}}$} & \multicolumn{1}{c}{$\sigma_{\Omega_K}$} & \multicolumn{1}{c}{$\sigma_{h}$} & \multicolumn{1}{c}{FOM} \\
\hline
\multirow{3}{*}{SKA1 (5,000 deg$^2$)}
 & 70  ~~~	&  0.258 & 0.972 & 0.00796 &  0.0476 &  0.00941 &  0.00943 &  28 \\
 & 150~~~	&  0.414 &  1.76  & 0.0115   &  0.0861 &  0.0109   &  0.0109   & 9  \\
 & 200~~~ & 0.582  &  2.66  & 0.0188   &  0.130   &  0.0122   &  0.0122   & 4 \\
\hline
\multirow{4}{*}{SKA2 (30,000 deg$^2$)}
  & 3.0~~~	&0.0328 & 0.116 & 0.00328 & 0.000158 & 0.00338 & 0.0034 & 547  \\
   &5.4~~~	&0.0407 & 0.137 & 0.00357 & 0.000169 & 0.00365 & 0.0040 & 426  \\
  & 23.0~~~ 	& 0.0912 & 0.322 & 0.00464 & 0.000224 & 0.00432& 0.0070 & 160  \\
\cline{2-9}
 & 0.0~~~	& 0.0273 & 0.100 & 0.00288 & 0.000148 & 0.00299 & 0.0029 & 699  \\
  \hline
\multirow{1}{*}{Euclid (15,000 deg$^2$)} & ---~~~ & 0.114 & 0.299 & 0.00907 & 0.000442 & 0.00944 & 0.0130 & 106  \\
\multirow{1}{*}{BOSS (10,000 deg$^2$)} & ---~~~ & 0.2416 & 0.9429 & 0.00810 & 0.0003958 & 0.0069091 & 0.01564 & 30  \\ 

\hline
\end{tabular} }
\caption{Forecast $1\sigma$ marginal errors and dark energy FOM for the various SKA1 and 2 reference experiments. A Euclid-like H$\alpha$ survey, BOSS forecasts, and a noise-free SKA2 configuration, are shown for comparison. (Planck CMB priors have been included in all cases.)}
\label{tab:marginals}\vspace{-1em}
\end{table*}


\section{Conclusions}

In this paper, we analysed the potential for producing precision cosmological constraints with future HI galaxy surveys using the SKA telescope. Neutral hydrogen (HI) is abundant in the late Universe, making it a prime candidate for detecting large numbers of galaxies which can then be used to trace the underlying dark matter distribution. In particular, modern radio receivers have the high sensitivity and bandwidth to detect the HI emission over an extremely wide redshift range, making it possible to trace the cosmological matter distribution over unprecedentedly large volumes.

Our analysis uses up-to-date simulations to calculate the expected galaxy number density and bias as a function of redshift and flux sensitivity. We have also provided a set of fitting formulas, Eqs. (\ref{equ:dndz}) and (\ref{bias}), that can be used to convert these results into number density and bias functions for specific experiments, such as the SKA or any other array.

One of our main conclusions is that although SKA1 will already detect a large number of HI galaxies, it will only be useful for cosmological applications up to $z\sim 0.7$ due to the sharp decline of the detected HI galaxy number density with redshift. This means that first, for a cosmological HI galaxy survey with SKA Phase 1, frequencies above $\sim 1$ GHz should be enough (i.e. Band 2). Moreover, these arrays will lack the sensitivity to detect enough galaxies to produce constraints that are competitive with contemporary optical and near-infrared galaxy surveys in the early 2020s.

On the other hand, the full SKA will push the HI galaxy detection limit up to $z \sim 2.0$ (requiring a larger band down to 500 MHz), and over the full visible sky, making it a prime cosmological survey instrument. Its sensitivity will allow us to produce an immense galaxy redshift survey over almost \sfrac{3}{4} of the sky, surpassing all other planned surveys in terms of precision measurements of the BAO. This should allow it to pin down the equation of state of dark energy with unprecedented precision. Note that, while we have concentrated on the BAO as the most robust large-scale structure observable, redshift space distortions and even the overall shape of the power spectrum contain a great deal of extra information that can also be used to constrain dark energy. In this sense, the forecasts in this paper represent conservative estimates of the cosmological constraints that can be achieved with the SKA (although recall that we optimistically neglected several nuisance parameters in our forecasts). 

Note that the SKA will also be able to produce competitive cosmological constraints using the HI intensity mapping (IM) technique \citep{Bull:2014rha}. IM surveys are sensitive to large-scale fluctuations in the HI brightness temperature, which can be used to recover information about the cosmological matter distribution without requiring high signal-to-noise detections of many individual sources. This means that the flux sensitivity of a telescope is used more efficiently in IM mode, as none of the detected signal need be discarded due to thresholding. Indeed, an IM survey with Phase 1 of the SKA will produce a dark energy figure of merit of at least half that of Euclid+BOSS \citep{Bull:2015nra}, in stark contrast to the underwhelming performance predicted for a Phase 1 HI galaxy survey (Section \ref{sec:comparison}).

This is not to say that galaxy surveys should be deprecated in favour of intensity mapping, however. Of the two, galaxy surveys are certainly the more tried-and-tested (and thus less risky) method -- the first large cosmological IM surveys are still a few years away, and a number of significant technical challenges (e.g. foreground contamination, polarisation leakage, autocorrelation calibration) remain to be solved \citep{Santos:2015gra}. In fact, a galaxy survey may be the preferred choice for a dark energy survey with Phase 2 of the SKA, as it will likely be easier to approach the sample variance limit for $z \lesssim 2$. This is because IM surveys are subject to a number of effective noise contributions separate from the instrumental noise (e.g. residual foregrounds and calibration errors) that are difficult to reduce to a negligible level, while galaxy surveys do not suffer from such residuals. Realistic simulations informed by experience with Phase 1 surveys will be needed to confirm this, however.

{\bf Acknowledgements:} SY thanks Pedro Ferreira for valuable guidance, and for hospitality during her visit to Oxford University where part of this work was done.
SY, MGS, RM and PO are supported by the South African Square Kilometre Array Project and  the National Research Foundation. PB is supported by European Research Council grant StG2010-257080. RM acknowledges support from the UK Science \& Technology Facilities Council (grant ST/K0090X/1). MS and MGS acknowledges support from FCT-Portugal under grant PTDC/FIS-AST/2194/2012.

\begin{table*}
\begin{centering}
{\renewcommand{\arraystretch}{1.3}
\begin{tabular}{rlllllllllll}
\hline 
\multirow{2}{*}{$S_{\rm rms}$ [$\mu$Jy]} & \multicolumn{11}{c}{Redshift bin} \\
 & \multicolumn{1}{c}{0.02} & \multicolumn{1}{c}{0.2} & \multicolumn{1}{c}{ 0.4} & \multicolumn{1}{c}{0.6} &  \multicolumn{1}{c}{0.8} & \multicolumn{1}{c}{1.0}   & \multicolumn{1}{c}{1.2}
& \multicolumn{1}{c}{1.4} & \multicolumn{1}{c}{ 1.6}& \multicolumn{1}{c}{ 1.8} & \multicolumn{1}{c}{ 2.0}   \\ 
\hline
 0~~ &	0.614 &	0.641 	&	0.678	 &	0.721	&  0.770	& 0.828 & 0.892 &0.963 &	1.04 & 	1.12 &1.21 \\
 1~~ & 	0.614 & 0.642	&	0.680	 &	0.738	&	0.836	& 0.965 & 1.13	&	1.34	&  1.59	&  1.82 &  2.14 \\
 3~~ &	0.614 & 0.643	&	0.695	 & 0.815	& 0.969 	& 1.14 	& 	1.34 &	1.59 & 	1.88 &	 2.15 &  2.53 \\
 5~~ &	0.614 &	0.644	& 	0.718	 & 0.868	& 1.04		& 1.22	& 1.44	& 	1.70	& 	2.02	& 	2.34	& 	2.71 \\
 6~~	& 	0.614 &	0.645 	& 0.730 	 & 0.886 	& 1.06		& 1.25	& 1.47 	&  1.73	& 	2.07 &	2.46	& 	2.86 \\
 7.3~~ &0.614 & 0.646	& 0.745		 & 0.907 	& 1.09		& 1.28	& 1.50	& 1.78	& 	2.12	& 	2.55	& 	2.86 \\
 10~~ & 0.614 & 0.650	& 0.770		 & 0.940	& 1.12		& 1.33	& 1.57	& 1.84	& 2.22	& 2.73	& 2.80 \\
 23~~ & 0.614 & 0.675	& 0.837		 & 1.021	& 1.22		& 1.45	& 1.75	& 1.95  &  --		&	--		&	--		\\
 40~~ & 0.614 & 0.706	& 0.879		 & 1.08		& 1.25		& 1.48	& 1.75	& 2.01 	&	--		& 	--		&	--		\\
 70~~ &	0.614 & 0.742	& 0.924		& 1.11		& 1.13		& 1.61	& 1.86 	&	-- 		&	-- 		& 	--   	& 	--	 	 \\
 100~~ &0.615 & 0.764	& 0.953		& 1.12		& 1.46		& 	-- 		& 	--		& 	--		& 	--		&	--		&	--		\\
 150~~ & 0.614& 0.787	& 0.982		& 1.11		&	-- 			&	--		&	--		&	--		&	--		&	--		&	--		\\
 200~~ & 0.614& 0.805	& 0.999		& 1.094		&	--			&	--		&	--		&	--		&	--		&	--		&	--		\\
\hline
\end{tabular} }
\renewcommand\thetable{A1}
\caption{Bias values calculated in each redshift bin of the simulation, as a function of flux rms.}
\end{centering}
\label{tab:bias}
\end{table*}

\balance

\bibliography{SKA_HI_galaxy_survey_v3.bib}

\end{document}